\documentstyle[12pt]{article}
\newcommand{\der}[2]{\frac{d #1}{d #2}}
\newcommand{\rau}{\partial}
\newcommand{\hb}[2]{\frac{\rau #1}{\rau #2}}
\newcommand{\da}{\alpha}
\newcommand{\de}{\varepsilon}
\newcommand{\bango}[2]{\begin{equation}#1  \label{#2} \end{equation}}
\newcommand{\nn}{\nonumber}
\newcommand{\naka}[1]{\begin{center} #1 \end{center}}

\newcommand{\enu}[1]{\begin{enumerate} #1 \end{enumerate}}
\newcommand{\lvec}[1]{\overrightarrow{#1}}
\newcommand{\llvec}[1]{\overleftarrow{#1}}
\newcommand{\braket}[2]{\langle #1 | #2 \rangle}
\newcommand{\eq}[1]{\begin{eqnarray} #1 \end{eqnarray}}
\newcommand{\ar}[4]{\left #1 \begin{array}{#2} #3 \end{array} \right#4}
\newcommand{\hyo}[2]{\begin{center}\begin{tabular}{#1}#2\end{tabular}\end{center}}

\newcommand{\D}{{\mathcal D}}
\newcommand{\K}{{\mathcal K}}

\newcommand{\itmz}[1]{\begin{itemize} #1 \end{itemize}}
\newcommand{\Bstar}{
\begin{picture}(10,10)
\put(5,2){\makebox(0,0){$\sqsubset$}}
\put(5,2){\makebox(0,0){$\star$}}
\end{picture}
}
\newcommand{\NAstar}{\diamond}
\newcommand{\Sstar}{
\begin{picture}(10,10)
\put(5,2){\makebox(0,0){$\Box$}}
\put(5,3){\makebox(0,0){$\star$}}
\end{picture}
}
\newcommand{\Ostar}{
\begin{picture}(12,10)
\put(6,3){\makebox(0,0){$\bigcirc$}}
\put(6,3){\makebox(0,0){$\star$}}
\end{picture}
}

\begin{document}
\naka{\Huge\bf Symmetrization of \\ the Berezin Star Product \\
and \\ Multiple Star Product Method}
\naka{\Large KAZUNORI WAKATSUKI
\footnote{waka@kiso.phys.metro-u.ac.jp}\footnote{waka@post.kek.jp}
}
\naka{\footnotemark[1]\it Department of Physics, Tokyo Metropolitan University, \\
Hachiohji, Tokyo, 192-0397, Japan}
\naka{\footnotemark[2]\it High Energy Accelerator Research Organization (KEK), \\
Tsukuba, Ibaraki 305-0801, Japan}

\naka{\bf Abstract}
\hspace{0.5cm}
We construct a multiple star product method and by using this method,
show that integral forms of some star products can be written in terms of
the path-integral.
This method can be applied to some examples. Especially, 
the associativity of the skew-symmetrized
Berezin star product proposed in \cite{SW}, is recovered
in large $N$ limit of the multiple star product. 
We also derive the path integral form of the Kontsevich star product
from the multiple Moyal star product.
This paper includes some reviews about star products.

\section{Introduction}
\hspace{0.5cm}
In recent years, a relation between superstring theory and a deformation
quantization has been explored.
D-branes, boundaries of open strings, are non-perturbative
object of superstring theory. Matrix Models \cite{BFSS,IKKT} were proposed as a D-brane action
a few years ago.
It is shown that a stable solution of this action is
non-commutative manifold \cite{CDS,AIIKKT}.
This non-commutativity, however, comes from the Moyal quantization.
So, this solution implies a flat D-brane\footnote{
We also derive a noncommutative gauge theory on a fuzzy sphere
from the matrix model\cite{IKTW}
}.
If we regard some D-brane as space-time,
we should study the deformation quantization of curved spaces
in order to realize the quantum gravity.

In order to proceed further it will be useful to clarify mathematical background of the deformation 
quantization. The star product was first introduced by Groenewold\cite{Gr},
which is now known as the Moyal product\cite{Mo}.
They associate an operator product to a noncommutative product of functions.
Here, the operators are mapped into the functions
by taking account the Weyl ordering.
The Weyl ordering means a skew-symmetric definition as we will see later.
Also Berezin tried to quantize curved phase spaces about 25 years ago
and succeeded to quantize some K\"{a}hler manifold e.g. sphere\cite{Be}.
Recently the Berezin quantization has been generalized to
arbitrary K\"{a}hler manifold\cite{RT}. However the Berezin quantization is
defined without skew symmetry, hence is not a generalization of Moyal one. 

The correlation between these methods of quantization is, however, not clear at all. In this paper,
we attempt to skew-symmetrize the Berezin quantization by means of the
multiple star product method.
The multiple star products reduce to the path-integral
form in large $N$ limit. (See \cite{Sh,Al} for the original ideas.)
As a result, our formulation turns similar to the path-integral form of the
Kontsevich quantization which is defined perturbatively 
on Poisson manifold\cite{Ko} but also can be described by a bosonic string path-integral\cite{CF}. 
Especially in the flat case, our star product coincides with the 
Kontsevich star product.

This paper consists of the following sections.
In section 2, we review the deformation quantization
e.g. Moyal \cite{Mo}, Berezin \cite{Be,MM} and
Kontsevich \cite{Ko,CF} quantization.
In section 3, we first construct the multiple star product method and
explain the symmetrized Berezin (or Wick type) star product\cite{SW,Ma}.
We also study its associativity in detail.
In section 4, we derive the path integral form of the Kontsevich
star product on the flat plane from the multiple star product method.
Section 5 is devoted to discussions.
Appendix includes some examples
that the multiple star product method is available.
\section{Deformation Quantization}
\hspace{0.5cm}
This section includes the definition and properties of 
the deformation quantization.
We also review Moyal, Berezin and Kontsevich quantization briefly as examples.

\subsection{General Definition and Property}
\hspace{0.5cm}
The deformation quantization\cite{BFFLS,St} is provided by a star product,
which is defined by
\bango{
f*g=\sum_{m=0}^{\infty}B_m(f,g)\lambda^m,
}{2.1}
where
\itmz{
\item $\lambda$ is a deformation parameter,
\item $f,g\in A=C^{\infty}(M)[\![\lambda]\!]$ \\
$C^{\infty}(M)[\![\lambda]\!]$ means that 
the coefficients of $\lambda$ power series are
$C^{\infty}$ functions on $M$ ,
\item $B_m$ are bi-differential operators ($B_m:A\times A\to A$),
}
The deformation quantization has the following properties:
\enu{
\item associativity
\bango{
f*(g*h)=(f*g)*h.
}{2.2}
\item $m=0$
\bango{
B_0(f,g)=fg.
}{2.2.5}
\item $m=1$
\bango{
B_1(f,g)-B_1(g,f)=\{f,g\}=2\sum_{i,j}\da^{ij}\rau_if\rau_jg,
}{2.3}
}
where $i,j=1,2,\cdots,d=dim(M)$ and
$\{\cdot,\cdot\}$ is a Poisson bracket which satisfies 
\bango{
\{f,\{g,h\}\}+\{g,\{h,f\}\}+\{h\{f,g\}\}=0,
}{2.3.4}
so the skew-symmetric bivector field $\da$ satisfies
\bango{
\da^{il}\rau_{l}\da^{jk}+\da^{jl}\rau_l\da^{ki}+\da^{kl}\rau_l\da^{ij}=0.
}{2.3.5}
It can be shown that one or more star product
determined by (\ref{2.2})(\ref{2.2.5})(\ref{2.3})
exist. The deformation quantization has the following equivalence
 called a gauge equivalence. 
$*$ and $*'$ are identified if 
\bango{
f'*'g'=D(f*g),
}{2.4}
where $f'=D(f),~g'=D(g)$ and $D$ is a differential operator ($D:A\to A$).
However, we can take two simple gauges. One is the
skew-symmetric gauge
\bango{
B_{1}(f,g)=\frac12\{ f,g\} =\sum_{i,j}\da^{ij}\rau_if\rau_jg.
}{2.4.3}
The other gauge is 
\bango{
B_{1}(f,g)=\sum_{i,j}\beta^{ij}\rau_if\rau_jg,
}{2.4.5}
where $\beta$ is the upper triangle matrix of $\da$ which satisfies $\da^{ij}
=\beta^{ij}-\beta^{ji}$ .
So we call the star product determined by (\ref{2.4.3}) and (\ref{2.4.5})
the skew-symmetric and asymmetric product respectively.
We have three concrete examples of deformation quantization,
which are put together in the following table.

\hyo{|c||c|c|c|}{
\hline
Manifold & flat plane & K\"{a}hler  & Poisson \\ \hline
Quantization & Moyal & Berezin & Kontsevich \\ \hline
Symbol & $\star$ & $\Bstar$ & $*$ \\ \hline
$m=1$ & skew-symmetric & asymmetric & skew-symmetric \\ \hline
}
We consider real two dimensional manifolds
for the sake of simplicity from now on.

\subsection{Moyal Quantization}
\hspace{0.5cm}
The Poisson bracket on the flat plane is defined by
\bango{
\{ f,g\} =\sum_{i,j}\de^{ij}\rau_if\rau_jg=\rau_xf\rau_pg-\rau_pf\rau_xg.
}{2.4.6}
Thus $\da^{ij}=\de^{ij}/2$ by (\ref{2.3}).
So we obtain the associative star product on the flat plane
i.e. the Moyal star product as
\eq{
f\star g(x,p)&=&f(x,p)e^{{\frac{\lambda}{2}} (\llvec{\rau_x}\lvec{\rau_{p}}
-\llvec{\rau_p}\lvec{\rau_{x}})}g(x,p)\nn \\
&=& fg+\lambda\frac12\{ f,g\} +O(\lambda^2) .
\label{2.5}}
This star product agrees with eq.(\ref{2.2.5}) and (\ref{2.4.3}),
and satisfies the associativity from the following results
\bango{
e_1\star (e_2\star e_3)=
e^{-\frac{\lambda}{2}(m_1n_2+m_2n_3+n_3m_1-n_1m_2-n_2m_3-m_3n_1)}
e_1e_2e_3=(e_1\star e_2)\star e_3,
}{2.5.5}
where $e_i$'s are the Fourier series
\[
e_i=e^{i(m_ix+n_ip)}.
\]
Here we require a usual canonical commutation relation
\bango{
[x,p]_{\star}=x\star p-p\star x=i\hbar,
}{2.6}
so that we obtain $\lambda=i\hbar$ .
Also this star product can be written by the integral form\cite{Ba},
because we have the following relations
\bango{
e^{i(mx+np)}\star e^{i(m'x+n'p)} =
e^{-\frac{i\hbar}{2}(mn'-m'n)}e^{i(mx+np)}e^{i(m'x+n'p)}
}{2.7.1}
and
\bango{
\int \frac{dw d\eta}{\pi\hbar}\frac{dw' d\eta'}{\pi\hbar}
e^{\frac{2i}{\hbar}S}e^{i(mw+n\eta)}e^{i(m'w'+n'\eta')}=
e^{-\frac{i\hbar}{2}(mn'-m'n)}e^{i(mx+np)}e^{i(m'x+n'p)},
}{2.7.2}
where
\[
S=
\ar{|}{ccc}{
1 &1 & 1 \\
x & w & w' \\
p & \eta & \eta'
}{|}.
\]
The left hand sides of eq.(\ref{2.7.1}) and (\ref{2.7.2}) are
equivalent, so that
we obtain the integral form of the Moyal star product
after multiplying arbitrary 
Fourier coefficients and integrating over $m,n$ as 
\bango{
f\star g(x,p)=\int \frac{dw d\eta}{\pi\hbar}\frac{dw' d\eta'}{\pi\hbar}
e^{\frac{2i}{\hbar}S}f(w,\eta)g(w',\eta').
}{2.8}

\subsection{Berezin Quantization}
\hspace{0.5cm}
The Poisson bracket on the K\"{a}hler manifold is given as
\bango{
\{ f,g\} =\frac{2}{i}h^{z\bar{z}}
(\rau_zf\rau_{\bar{z}}g-\rau_{\bar{z}}f\rau_zg),
}{2.8.1}
where $h^{z\bar{z}}$ is the inverse of a K\"{a}hler metric
\[
h_{z\bar{z}}=\rau_z\rau_{\bar{z}}K(z,\bar{z}),
\]
and $K(z,\bar{z})$ is a K\"{a}hler potential.
The factor $2/i$ in eq.(\ref{2.8.1}) is necessary in order that
the Poisson bracket becomes eq.(\ref{2.4.6}) in the flat case.
The original Berezin quantization covers only Ricci flat K\"{a}hler manifold.
The Berezin star product is defined by
\bango{
f\Bstar g(z,\bar{z})=
\int d\mu_{\nu}(v,\bar{v}) e^{\frac{1}{\nu}
\Phi(z,\bar{z},v,\bar{v})}f(z,\bar{v})g(v,\bar{z}),
}{2.9}
where $\Phi(z,\bar{z},v,\bar{v})$ is called the Calabi function and
defined by the K\"{a}hler potential as
\bango{
\Phi(z,\bar{z},v,\bar{v})=
K(z,\bar{v})+K(v,\bar{z})-K(z,\bar{z})-K(v,\bar{v}),
}{2.10}
and the measure $d\mu_{\nu}$ is determined by the metric as
\bango{
d\mu_{\nu} (z,\bar{z})=h_{z\bar{z}}\frac{idz\wedge d\bar{z}}{2\pi\nu}.
}{2.10.1}
This star product can be expanded around $\lambda=\nu/2i=0$ as follows
\bango{
f\Bstar g(z,\bar{z})=fg+\lambda\Big(B_1^{+}(f,g)+B_1^{-}(f,g)\Big)
+O(\lambda^2),
}{2.10.2}
where $B_1^{+}$ and $B_1^{-}$ are a symmetric part and a skew-symmetric part
respectively as
\bango{
B_1^{+}=2iAfg-\frac12\{ f,g\}_+~,~
B_1^{-}=\frac12\{ f,g\}
}{2.10.3}
and
\[
A=\frac12 h^{z\bar{z}}\rau_z\rau_{\bar{z}}\log{h_{z\bar{z}}}~,~
\{ f,g\}_+ =\frac{2}{i}h^{z\bar{z}}
(\rau_zf\rau_{\bar{z}}g+\rau_{\bar{z}}f\rau_zg).
\]
If the manifold $M$ is the K\"{a}hler manifold, $A=0$ \cite{RT}. 
Thus (\ref{2.2.5}) and (\ref{2.3}) are satisfied.
Also the associativity is shown as the following.
\[
\Big((f\Bstar g)\Bstar h\Big)(z,\bar{z})
=
\int d\mu_{\nu}(v,\bar{v})d\mu_{\nu}
(u,\bar{u})f(z,\bar{v})g(v,\bar{u})h(u,\bar{z})
e^{\frac{1}{\nu}
\big(\Phi(z,\bar{u},v,\bar{v})+\Phi(z,\bar{z},u,\bar{u})\big)}.
\]
\[
\Big(f\Bstar(g\Bstar h)\Big)(z,\bar{z})=
\int d\mu_{\nu}(v,\bar{v})d\mu_{\nu}(u,\bar{u})f(z,\bar{v})g(v,\bar{u})h(u,\bar{z})
e^{\frac{1}{\nu}\big(
\Phi(v,\bar{z},u,\bar{u})+\Phi(z,\bar{z},v,\bar{v})\big)}.
\]
The Calabi function clearly satisfies 
\[
\Phi(z,\bar{u},v,\bar{v})+\Phi(z,\bar{z},u,\bar{u})=
\Phi(v,\bar{z},u,\bar{u})+\Phi(z,\bar{z},v,\bar{v}),
\]
so the associativity is shown:
\bango{
\Big((f\Bstar g)\Bstar h\Big)(z,\bar{z})
=\Big(f\Bstar(g\Bstar h)\Big)(z,\bar{z}).
}{2.12}

\subsection{Kontsevich Quantization}
\hspace{0.5cm}
The Kontsevich quantization covers the Poisson manifold ($M$) 
which is a general manifold with the Poisson structure.
He perturbatively solved $B_m(f,g)$'s under the conditions
(\ref{2.2}),(\ref{2.2.5}) and (\ref{2.4.3}) as
\bango{
B_m(f,g)=\sum_{\Gamma\in G_m}w_{\Gamma}B_{\Gamma}(f,g),
}{bm}
where $G_m$ is a set of diagrams related to the number $m$ ,
$B_{\Gamma}(f,g)$ is a bi-differential operator
determined by the Feynman diagram and
$\omega_{\Gamma}$ is a weight
\cite{Ko}.
Thus Kontsevich defines the star product on the Poisson manifold by a
formal power series of $\lambda$ as
\bango{
f*g=\sum_{m=0}^{\infty}\lambda^m\sum_{\Gamma\in G_m}w_{\Gamma}B_{\Gamma}(f,g).
}{2.20}
Also Cattaneo and Felder have shown that the
Kontsevich star product (\ref{2.20}) coincides with the path integral form
of a topological bosonic string (non-linear sigma model):
\bango{
f*g(x)=\int_{X(\infty)=x}f(X(1)g(X(0))e^{\frac{i}{\hbar}S[X,\eta]}
\D X \D \eta, 
}{2.23}
where the action is defined on a disk $D$ as
\[
S[X,\eta]=\int_D \eta_i(u)\wedge dX^i(u)+\frac12\da^{ij}(X(u))
\eta_i(u)\wedge \eta_j(u),
\]
and
\itmz{
\item $D=\{u\in R^2~,~|u|\leq 1\}$,
\item $X$ and $\eta$ are real bosonic fields,
\item $X : D\to M$,
\item $\eta$ is a differential 1-form on $D$ : $X^{*}(T^{*}M)\otimes T^{*}D$.
}
In the symplectic case, the action can be integrated over $\eta$ and becomes
a boundary integration by the Stokes's theorem as
\bango{
f*_{\mathrm{symp}}g(x)=\int_{\gamma(\pm\infty)=x}f(\gamma(1))g(\gamma(0))
e^{\frac{i}{\hbar}\int_{\gamma}d^{-1}\omega}d\gamma,
}{2.26}
where $\gamma$ is a loop trajectory from $x$ to $x$ .


\section{Symmetrized Berezin star product}
\hspace{0.5cm}
In this section,
we first explain the multiple star product method.
Next we define the S-star product
to clarify the relationship between Moyal and Berezin quantization.
However the S-star product is not associative in the curved space. So using 
the multiple star product method, we derive an associative star product i.e.
the O-star product.

\subsection{Multiple star product method}
Generally the integral forms of the star products can be written as
\bango{
f\NAstar g(\da)=\int 
d\mu_{\lambda}(\beta,\gamma)~e^{\K_{\lambda}(\da,\beta,\gamma)}f(\beta)g(\gamma),}
{gif}
where $\da,\beta,\gamma\in M$, $\K_{\lambda}=\K/\lambda$ is an integral kernel
and $d\mu_{\lambda}=d\mu/\lambda^2$ is a measure which relates two points on $M$.
We assume that this star product $\NAstar$ satisfies the followings:
\bango{
f\NAstar g=fg+\lambda\frac{\{ f,g\}}{2}+O(\lambda^2),
}{gif0}
\bango{
f\NAstar 1=1\NAstar f=f,
}{gif1}
\bango{
d\mu_{\lambda}(\beta,\gamma)=d\mu_{\lambda}(\gamma,\beta).
}{gif2}
We also add a assumption $\K_{\lambda}(\da,\beta,\beta)=0$ in particular.
Note that we don't require this star product $\NAstar$ is associative.
Then we call this product $\NAstar$ the \textit{non-associative} star product.

Next we define the multiple star product of $\NAstar$ as
\bango{
A^N(f)=f_{N/N}\NAstar  f_{N-1/N}\NAstar  \cdots \NAstar  f_{2/N}\NAstar  f_{1/N}.
}{msp}
An equivalence of the forward product $\llvec{A}^N$ and the backward product
$\lvec{A}^N$ is necessary at least in order that
$A^N$ is well-defined where
\eq{
\llvec{A}^N(f)&:=&
(f_{N/N}\NAstar (f_{N-1/N}\NAstar (\cdots\NAstar (f_{1/N}\NAstar  1)\cdots))) \nn \\
&=& \int \Big(\prod_{i=1}^{N}d\mu_{\lambda N}(\beta_{i/N},\gamma_{i/N})f_{i/N}(\beta_{i/N})\Big)
\exp{\sum_{i=1}^{N}\frac1N\K_{\lambda}(\gamma_{i+1/N},\beta_{i/N},\gamma_{i/N})},
\label{fp}}
\eq{
\lvec{A}^N(f)&:=&
(((\cdots(1\NAstar  f_{N/N})\NAstar \cdots)\NAstar  f_{2/N})\NAstar  f_{1/N}) \nn \\
&=& \int \Big(\prod_{i=1}^{N}d\mu_{\lambda N}(\beta_{i/N},\gamma_{i/N})f_{i/N}(\beta_{i/N})\Big)
\exp{\sum_{i=1}^{N}\frac1N\K_{\lambda}(\gamma_{i-1/N},\gamma_{i/N},\beta_{i/N})},
\label{bp}}
\bango{
\da=\beta_0=\beta_{N+1/N}=\gamma_0=\gamma_{N+1/N}.
}{bc}
Note that we change the deformation parameter $\lambda$ to $\lambda N$.
>From this equivalence, we obtain a condition
\bango{
\frac1N\sum_{i=1}^N\Big( \K_{\lambda}(\gamma_{i+1/N},\beta_{i/N},\gamma_{i/N})
-\K_{\lambda}(\gamma_{i-1/N},\gamma_{i/N},\beta_{i/N})\Big)=0.
}{equiv}
Using the boundary condition (\ref{bc}) and the additional condition
$\K(\da,\beta,\beta)=0$, this condition (\ref{equiv}) is also deformed as
\bango{
\frac1N\sum_{i=0}^N \Big(\K_{\lambda}(\gamma_{i+1/N},\beta_{i/N},\gamma_{i/N})
-\K_{\lambda}(\gamma_{i/N},\gamma_{i+1/N},\beta_{i+1/N})\Big)=0.
}{dequiv}
This condition corresponds to the associativity condition
in the case of $f_{i/N}=1$ except for three $f_{i/N}$'s.
We denote $A^N(f):=\llvec{A}^N(f)=\lvec{A}^N(f)$ 
when $\K_{\lambda}$ satisfies eq.(\ref{dequiv}).

\subsection{Relationship between Moyal and Berezin Star Product}
\hspace{0.5cm}
The Berezin star product in the flat case, coincides with the Moyal's one
except for skew-symmetry or asymmetry. This difference is 
explained as follows. First, we write the Moyal star product in complex
variables to make clear the correspondence to Berezin star product,
\bango{
f\star g(z,\bar{z})
= f(z,\bar{z})e^{\hbar(\llvec{\rau_z}\lvec{\rau_{\bar{z}}}
-\llvec{\rau_{\bar{z}}}\lvec{\rau_z})}g(z,\bar{z}),
}{2.13}
where $z=x+ip$.
This star product is gauge equivalent(\ref{2.4})
to $\star_{st}$ and $\star_{ar}$ where
\bango{
\star_{st}=e^{2\hbar\llvec{\rau_z}\lvec{\rau_{\bar{z}}}}~~~ {\mathrm{and}}~~~
\star_{ar}=e^{-2\llvec{\rau_{\bar{z}}}\lvec{\rau_z}},
}{standar}
because the gauge equivalent condition is satisfied in the case of
\bango{
D=e^{\hbar\llvec{\rau_z}\lvec{\rau_{\bar{z}}}}~~~ {\mathrm{and}}~~~
D=e^{-\hbar\llvec{\rau_{\bar{z}}}\lvec{\rau_z}},
}{gaugetra}
respectively\cite{Vo,Be,APS}. Thus we obtain a star product relation,
\bango{
\star=(\star_{st}\star_{ar})^\frac12.
}{starrelation}
Here $\star_{st}^{\frac12}$ and $\star_{ar}^{\frac12}$ 
can be written in the integral forms\cite{APS} as
\eq{
f\star_{st}^{\frac12}g(z,\bar{z})&=&\int\frac{idw\wedge d\bar{w}}{2\pi\theta}
e^{\frac{1}{\theta}|w-z|^2}f(w,\bar{z})g(z,\bar{w}), \nn \\
f\star_{ar}^{\frac12}g(z,\bar{z})&=&\int\frac{idv\wedge d\bar{v}}{2\pi(-\theta)}
e^{-\frac{1}{\theta}|v-z|^2}f(z,\bar{v})g(v,\bar{z}),
\label{intforms}}
where $\theta=-\hbar$. Thus we obtain
\bango{
f\star g(z,\bar{z})=-\int \frac{idv\wedge d\bar{v}}{2\pi\theta}
\frac{idw\wedge d\bar{w}}{2\pi\theta}
e^{\frac{1}{\theta}(-|v-z|^2+|w-z|^2)}f(w,\bar{v})g(v,\bar{w}).
}{complexmoyalstar}
The star products (\ref{intforms}) are two types of the Berezin star product
on the flat plane i.e. the term $-|v-z|^2$ is the Calabi function on the
flat plane. Taking this result into account, we generalize the Moyal
star product $\star$ to
the S-star product on the Ricci flat K\"{a}hler manifold
\footnote{In \cite{Ma}, it is discussed that
the S-star product may be available to
general K\"{a}hler manifold.}, which is defined by
\bango{
f\Sstar g(z,\bar{z}):=\int d\mu_{\theta}(v,\bar{v})d\mu_{-\theta}(w,\bar{w})
\exp{\frac{1}{\theta}\Big(\Phi(z,\bar{z};v,\bar{v})-
\Phi(z,\bar{z};w,\bar{w})\Big)}f(w,\bar{v})g(v,\bar{w}),
}{Sstar}
where $d\mu_{\theta}(z,\bar{z})=h_{z\bar{z}}idz\wedge d\bar{z}/2\pi\theta$ 
similarly to the definition (\ref{2.10.1}).
However, this star product is not associative unless flat.
This complication is overcome by using the multiple star product method.

\subsection{Associativity of Symmetrized Berezin Star Product}
\hspace{0.5cm}
In this section, we attempt to recover the associativity of the S-star product.
First, we show that the S-star product is non-associative star product.
In the case of the S-star product, we know the following correspondence:
\bango{
\da=(z,\bar{z}),~\beta=(w,\bar{v}),~\gamma=(v,\bar{w}),
}{corresp1}
\bango{
\lambda=\theta,~d\mu_{\lambda}(\beta,\gamma)=d\mu_{\theta}(v,\bar{v})
d\mu_{-\theta}(w,\bar{w}),
}{corresp2}
\bango{
\K_{\lambda}(\da,\beta,\gamma)=\frac{1}{\theta}\Big(
\Phi(z,\bar{z};v,\bar{v})-\Phi(z,\bar{z};w,\bar{w})\Big).
}{corresp3}
Thus the S-star product clearly satisfies the conditions (\ref{gif2}) and 
$\K_{\lambda}(\da,\beta,\beta)=0$ . Also it is 
derived in \cite{Ma} that
this product satisfies the conditions (\ref{gif0}) and (\ref{gif1}).

\begin{figure}
\begin{picture}(200,250)
\put(50,240){\makebox(0,0){$f_3\Ostar (f_2\Ostar f_1)$}}
\put(17,180){\makebox(0,0){$=$}}
\put(50,180){\circle{40}}
\put(50,200){\circle{4}}
\put(36,166){\circle*{4}}
\put(25,160){\makebox(0,0){$f_3$}}
\put(64,166){\circle*{4}}
\put(90,160){\makebox(0,0){$f_2\Ostar f_1$}}
\put(17,120){\makebox(0,0){$=$}}
\put(50,120){\circle{40}}
\put(50,140){\circle{4}}
\put(36,106){\circle*{4}}
\put(30,100){\makebox(0,0){$f_3$}}
\put(64,106){\circle{4}}
\put(78,92){\circle{40}}
\put(98,92){\circle*{4}}
\put(108,92){\makebox(0,0){$f_1$}}
\put(78,72){\circle*{4}}
\put(78,62){\makebox(0,0){$f_2$}}
\put(17,30){\makebox(0,0){$=$}}
\put(70,30){\circle{40}}
\put(70,50){\circle{4}}
\put(70,10){\circle*{4}}
\put(70,0){\makebox(0,0){$f_2$}}
\put(50,30){\circle*{4}}
\put(40,30){\makebox(0,0){$f_3$}}
\put(90,30){\circle*{4}}
\put(100,30){\makebox(0,0){$f_1$}}
\put(200,240){\makebox(0,0){$(f_3\Ostar f_2)\Ostar f_1$}}
\put(167,180){\makebox(0,0){$=$}}
\put(228,180){\circle{40}}
\put(228,200){\circle{4}}
\put(214,166){\circle*{4}}
\put(188,160){\makebox(0,0){$f_3\Ostar f_2$}}
\put(242,166){\circle*{4}}
\put(253,160){\makebox(0,0){$f_1$}}
\put(167,120){\makebox(0,0){$=$}}
\put(228,120){\circle{40}}
\put(228,140){\circle{4}}
\put(242,106){\circle*{4}}
\put(250,100){\makebox(0,0){$f_1$}}
\put(214,106){\circle{4}}
\put(200,92){\circle{40}}
\put(180,92){\circle*{4}}
\put(170,92){\makebox(0,0){$f_3$}}
\put(200,72){\circle*{4}}
\put(200,62){\makebox(0,0){$f_2$}}
\put(167,30){\makebox(0,0){$=$}}
\put(220,30){\circle{40}}
\put(220,50){\circle{4}}
\put(220,10){\circle*{4}}
\put(220,0){\makebox(0,0){$f_2$}}
\put(200,30){\circle*{4}}
\put(190,30){\makebox(0,0){$f_3$}}
\put(240,30){\circle*{4}}
\put(250,30){\makebox(0,0){$f_1$}}
\put(140,0){\line(0,1){250}}
\end{picture}
\caption[associativity]{If we denote O-star product by a circle,
the associativity is shown as above. Here $\circ$ point means
the boundary of O-star product as $(z,\bar{z})$ .}
\label{associativity}
\end{figure}
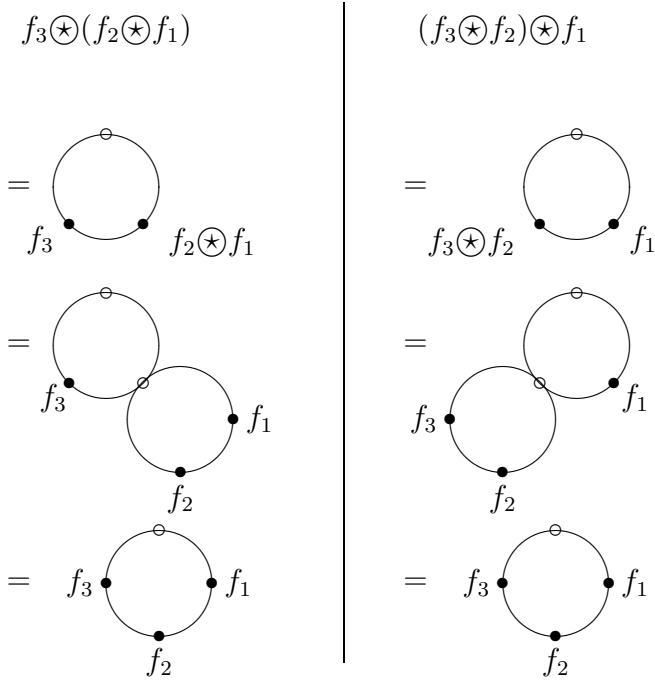

Next, we survey whether the S-star product satisfies the condition (\ref{dequiv}) 
or not. In the S-star product, $lhs$ of (\ref{dequiv}) is written in terms of
the K\"{a}hler potential $K$ as
\eq{
lhs&=&\frac{1}{\theta}\frac1N\sum_{i=0}^{N}\Big(K(v_{i+1/N},\bar{v}_{i/N})
+K(v_{i/N},\bar{v}_{i+1/N})-2K(v_{i/N},\bar{v}_{i/N})\Big) \nn \\
&&-\Big(
K(w_{i+1/N},\bar{w}_{i/N})+K(w_{i/N},\bar{w}_{i+1/N})-2K(w_{i/N},\bar{w}_{i/N})
\Big).
\label{Kcond}}
This result is non-zero but becomes zero in the large $N$ limit as
\bango{
lhs\to\frac{1}{\theta}\int_{0}^{1}d\tau~
d\Big(K(v,\bar{v})-K(w,\bar{w})\Big)=0,
}{total0}
where $v,w=v(\tau),w(\tau)$ 
and the boundary conditions (\ref{bc}) become
\bango{
v(0)=v(1)=w(0)=w(1)=z,~\bar{v}(0)=\bar{v}(1)
=\bar{w}(0)=\bar{w}(1)=\bar{z}.
}{bc1}
Thus $A^N(f)$ is ill-defined but $A(f)=\lim_{N\to\infty}A^N(f)$
is well-defined. Then we call this star product \textit
{pseudo-associative} product.
By using $A(f)$ and
\bango{
f_{i/N}=
\ar{\{ }{lll}{
g & (i/N=i_1/N\to\tau_1) \\
f & (i/N=i_2/N\to\tau_2) \\
1 & (i/N\to\tau \ne \tau_1,\tau_2)
}{.},
}{f}
we construct an associative star product $\Ostar$ in
terms of the path integral form as
\eq{
&&f\Ostar g(z,\bar{z}) := A(f) \nn \\
&=& \cdots 1 \Sstar  1 \Sstar  f \Sstar  1
\Sstar  1 \cdots 1 \Sstar  1 \Sstar  g \Sstar  1 \Sstar  1 \cdots \nn \\
&=& \lim_{N\to\infty}\int\prod_{i=1}^Nd\mu_{\theta N}(v_{i/N},\bar{v}_{i/N})
d\mu_{-\theta N}(w_{i/N},\bar{w}_{i/N})\exp{\Big[\frac{i}{\theta}
S_i\Big]}f(v_{i_2/N},\bar{w}_{i_2/N})g(v_{i_1/N},\bar{w}_{i_1/N}) \nn \\
&=& \int {\mathcal D}\mu_{\theta}(v,\bar{v}) {\mathcal D}\mu_{-\theta}
(w,\bar{w})
\exp{\Big[\frac{i}{\theta}S\Big]} f\Big(v(\tau_2),\bar{w}(\tau_2)\Big)
g\Big(v(\tau_1),\bar{w}(\tau_1)\Big),
\label{4.13}}
where actions are written as follows
\bango{
S_i=\frac1i\frac1N\sum_{i=1}^{N}\Big(
\Phi(v_{i+1/N},\bar{w}_{i+1/N};v_{i/N},\bar{v}_{i/N})
-\Phi(v_{i+1/N},\bar{w}_{i+1/N};w_{i/N},\bar{w}_{i/N})\Big),
}{Si}
\bango{
S=
\frac{1}{i}\int_{-\infty}^{\infty} \bigg[
\Big(\psi(v,\bar{v})-\psi(v,\bar{w})\Big)\hb{v}{\tau}
-\Big(\bar{\psi}(w,\bar{w})-\bar{\psi}(v,\bar{w})\Big)
\hb{\bar{v}}{\tau} \bigg]d\tau,
}{S}
and the path integral measure is defined by
\bango{
\D\mu_{\theta}(v,\bar{v}):=
\lim_{N\to\infty}\prod_{i=1}^{N}d\mu_{\theta N}(v_{i/N},\bar{v}_{i/N}).
}{pathmeasure}
Note that $\psi(z,\bar{z})$ is a canonical conjugation of $z$,
which is defined by
\bango{
\psi(z,\bar{z}):=\hb{K(z,\bar{z})}{z}.
}{psi}
As above, associative symmetrized Berezin star product is
defined as O-star product correctly. 
The associativity is satisfied as illustrated in Figure \ref{associativity}.

\section{Construction of Kontsevich Star Product from Multiple Star
Product Method}
\hspace{0.5cm}
In this section, we show that the multiple Moyal star product
corresponds to the path integral form of the Kontsevich star product
on the flat plane.
First preparatory to this derivation, we
write the multiple Moyal star product in large $N$ limit as
\eq{
A_{\star}(f) &:=&\lim_{N\to\infty}f_{N/N}(x,p)\star  f_{N-1/N}(x,p)\star \cdots
\star  f_{2/N}(x,p) \star  f_{1/N}(x,p) \nn \\
&=& \lim_{N\to\infty}\int\prod_{i=1}^{N}\frac{d\xi_{i/N}d\eta_{i/N}}{\pi\hbar N}
\frac{d\xi'_{i/N}d\eta'_{i/N}}{\pi\hbar N}f_{i/N}(\xi_{i/N},\eta_{i/N})
\exp{\Big[\frac{2i}{\hbar N}\sum_{i=1}^{N}
\ar{|}{ccc}{
1 & 1 & 1 \\
\xi'_{i+1/N} & \xi_{i/N} & \xi'_{i/N} \\
\eta'_{i+1/N} & \eta_{i/N} & \eta'_{i/N}
}{|}\Big]} \nn \\
&=& \int \D\xi \D\eta \D\xi' \D\eta'
\prod_{\tau=0}^1f(\tau;\xi,\eta)
\exp{\frac{2i}{\hbar}\int_{0}^{1}d\tau
\Big[ \der{\xi'}{\tau}(\eta-\eta')-(\xi-\xi')\der{\eta'}{\tau}\Big]},
\label{5.1}}
where real fields $\xi, \eta, \xi', \eta'$ have boundary conditions
\bango{
x=\xi(0)=\xi(1)=\xi'(0)=\xi'(1)~,~p=\eta(0)=\eta(1)=\eta'(0)=\eta'(1),
}{bcxp}
and functional measures are defined as follows
\bango{
\D\xi:=\lim_{N\to\infty}\prod_{i=1}^N\frac{d\xi_{i/N}}{\pi\hbar N},~~
\cdots etc.
}{measures}
In eq.(\ref{5.1}),
we can integrate out $\xi',\eta'$ by using partial integration and obtain
simplified form 
\bango{
A_{\star}(f)=\int\D\xi\D\eta \prod_{\tau=0}^{1}f_{\tau}(\xi(\tau),\eta(\tau))
\exp{\frac{i}{\hbar}\int_{0}^{1}
\eta\hb{\xi}{\tau}d\tau}.
}{5.2}
Here we can change the integration area of $\tau$ $(0,1)$ to
$(-\infty,\infty)$ by a reparametrization of $\tau$ .
Thus eq.(\ref{5.2}) and the boundary conditions (\ref{bcxp}) are changed as
\bango{
A_{\star}(f)=\int\D\xi\D\eta \prod_{\tau=-\infty}^{\infty}
f_{\tau}(\xi(\tau),\eta(\tau))
\exp{\frac{i}{\hbar}\int_{-\infty}^{\infty}
\eta\hb{\xi}{\tau}d\tau}.
}{Astar}
\bango{
x=\xi(\pm\infty)=\xi'(\pm\infty),~p=\eta(\pm\infty)=\eta'(\pm\infty).
}{bcxp2}

Next by using eq.(\ref{Astar}), we show that 
the multiple Moyal star product is included in the path integral form
of the Kontsevich star product(\ref{2.26}). If we put
\bango{
f_{\tau}(x,p)=
\ar{\{}{ll}{
f(x,p)&(\tau=1)\\
g(x,p)&(\tau=0)\\
1&(\tau\ne 0,1)
}{.},
}{ftauK}
then
\eq{
f\star g(x,p)&=&A_{\star}(f) \nn \\
&=& \lim_{N\to\infty}\cdots \star  1\star  f(x,p)\star  1\star  
\cdots\star  1 \star  g(x,p)\star  1\star  \cdots \nn \\
&=& \int {\mathcal D}\xi{\mathcal D}\eta ~f(\xi(1),\eta(1))
g(\xi(0),\eta(0))\exp{\frac{i}{\hbar}\int_\gamma
d^{-1}\omega_0},
\label{5.5}}
where 
\bango{
d^{-1}\omega_0:=(d\xi)\eta=\eta\der{\xi}{\tau}d\tau,
}{dinvomega0}
and
\bango{
\omega_0=d(d^{-1}\omega_0)=d\xi\wedge d\eta.
}{omega0}
Eq.(\ref{5.5}) corresponds to eq.(\ref{2.26}) on the flat plane.

\section{Discussions}
\hspace{0.5cm}
In this paper, we have proposed the multiple star product method.
It is useful for constructing associative star products from
\textit{pseudo-associative} star products.
We have shown that a \textit{pseudo-associative} S-star product
becomes an associative O-star product by using the multiple star product method.
It is explained as follows. Although the \textit{pseudo-associative} products break associativity 
condition a little, the multiple star product method, which is a set of
infinite \textit{pseudo-associative} product, 
smoothes and overcomes this risk and the associativity condition
(\ref{dequiv}) is satisfied.
In consequence, the \textit{pseudo-associative} product turns to an associative product within 
the frame work of the path integral formalism.

The multiple star product method also has been available
to well-known associative products e.g. the Moyal star product.
The multiple Moyal star product coincides with the path integral form
of the Kontsevich star product on the flat plane.
This result implies a justice of the multiple star product method.

\appendix
\section{Other Examples}
\hspace{0.5cm}
By using eq.(\ref{Astar}), we can obtain the
transition amplitude in quantum dynamics and the bosonic string generating
function. If we put in eq.(\ref{Astar})
\bango{
f_\tau(x,p)=
\ar{\{}{ll}{
\psi_I(x) & (\tau=t_I) \\
\bar{\psi_F}(x) & (\tau=t_F) \\
e^{-\frac{i}{\hbar}H(x,p)} & (t_I<\tau<t_F) \\
1 & (\tau<t_I~,~t_F<\tau)
}{.},
}{ta}
we obtain
\eq{
A_{\star}(f)&=&\lim_{N\to\infty}\cdots \star  1\star  \bar{\psi}_F(x)\star  
e^{-\frac{i\de}{\hbar}H(x,p)}\star  \cdots \star 
e^{-\frac{i\de}{\hbar}H(x,p)}\star  \psi_I(x) \star  1\star \cdots \nn \\
&=& \int {\mathcal D}\xi {\mathcal D}\eta ~
\bar{\psi}_F(\xi(t_F))\psi_I(\xi(t_I))e^{\frac{i}{\hbar}\int_{t_I}^{t_F} 
\big(\eta\hb{\xi}{\tau}-
H(\xi,\eta)\big)d\tau} \nn \\
&=& 
\braket{\psi_F,t_F}{\psi_I,t_I},
}
where integration of $\tau<t_I, t_F<\tau$ vanishes
because of $f(\tau;x,p)=1$ .

Next, 
a bosonic string generating function can be derived from infinite dimensional
multiple Moyal star products. In (\ref{Astar}), we put
\bango{
f_{\tau}(x,p)=e^{i\big(k(\tau)x+m(\tau)p\big)},
}{bsgf}
and obtain
\eq{
A_{\star}(f)
&=& \lim_{N\to\infty}e^{i\big(k(\infty)x+m(\infty)p\big)}\star \cdots
\star e^{i\big(k(0)x+m(0)p\big)}\star\cdots
\star e^{i\big(k(-\infty)x+m(-\infty)p\big)} \nn \\
&=&
\int {\mathcal D}\xi {\mathcal D}\eta~
e^{\frac{i}{\hbar}\int_{-\infty}^{\infty} \big(
\eta\hb{\xi}{\tau}+\hbar(k\xi+m\eta) \big)d\tau}.
}
Here, we generalize 
$x\to x_{n,\mu}~,~p\to p_{n}^{\mu}$ where $n$ runs from $0$ to $\infty$ and
$\mu$ runs from $1$ to dimension $d$ . $n$ can be changed to continuous
parameter $\sigma$ by the following definitions and relations
\[
x_{\mu}(\sigma):=\sum_{n=0}^{\infty}x_{n,\mu}\cos{n\sigma}~,~
p^{\mu}(\sigma):=\sum_{n=0}^{\infty}\frac{p_n^{\mu}}{n}\sin{n\sigma},
\]
\[
\sum_{n=0}^{\infty}x_{n,\mu}p_n^{\mu}=
2\int_0^{2\pi}x_{\mu}\hb{p^{\mu}}{\sigma}d\sigma.
\]
Substituting above relation into $A_{mn}$ , we obtain
\eq{
A_{\star}(f)=
\int {\mathcal D}X
\exp{\Bigg[\frac{2i}{\hbar}\int_{-\infty}^{\infty}d\tau\int_0^{2\pi}
 d\sigma \Bigg(
\hb{X^{\mu}}{\sigma}\hb{X_{\mu}}{\tau}
+J^{\mu}X_{\mu}\Bigg)\Bigg]},
}
\[
X_{\mu}=\frac{\xi_{\mu}+\eta_{\mu}}{\sqrt{2}}~,~
J^{\mu}=\hbar\Big(\hb{n^{\mu}}{\sigma}-\hb{m^{\mu}}{\sigma}\Big)~,~
\D X=\D \xi \D \eta.
\]
$A_{\star}(f)$ becomes the bosonic string generating function.
More details are in \cite{SW}.

\textbf{Acknowledgments}\\
I'm indebted to Professors Y.Maeda and S.Saito for useful discussions.

\end{document}